\documentclass[twocolumn,twocolumn]{IEEEtran}
\usepackage[T1]{fontenc}
\usepackage{color}
\usepackage{float}
\usepackage{mathrsfs}
\usepackage{amsmath}
\usepackage{amssymb}
\usepackage{graphicx}
\usepackage[unicode=true,
 bookmarks=true,bookmarksnumbered=true,bookmarksopen=true,bookmarksopenlevel=1,
 breaklinks=false,pdfborder={0 0 0},pdfborderstyle={},backref=false,colorlinks=false]
 {hyperref}
\hypersetup{pdftitle={Your Title},
 pdfauthor={ZhuYian},
 pdfpagelayout=OneColumn, pdfnewwindow=true, pdfstartview=XYZ, plainpages=false}

\makeatletter

\floatstyle{ruled}
\newfloat{algorithm}{tbp}{loa}
\providecommand{\algorithmname}{Algorithm}
\floatname{algorithm}{\protect\algorithmname}

\usepackage[caption=false,font=footnotesize]{subfig}
\usepackage{algorithm}
\usepackage{algorithmic}

\usepackage{multirow} %multirow for format of table 
\usepackage{amsmath} 
\usepackage{xcolor}

\allowdisplaybreaks[4]

\ifCLASSOPTIONcompsoc
\usepackage[caption=false,font=normalsize,labelfont=sf,textfont=sf]{subfig}
\else
\usepackage[caption=false,font=footnotesize]{subfig}
\fi

\usepackage{cite}
\usepackage{bm}
\usepackage{algorithmic}
\usepackage{algorithm}
\usepackage{graphicx}
\IEEEoverridecommandlockouts

\usepackage{lettrine}

\usepackage{geometry}
\@ifundefined{showcaptionsetup}{}{%
 \PassOptionsToPackage{caption=false}{subfig}}
\usepackage{subfig}
\makeatother
\begin{document}

\title{\textcolor{black}{UAV Trajectory Tracking via RNN-enhanced IMM-KF with ADS-B Data}}

\author{\IEEEauthorblockN{Yian Zhu$^{\ast}$, Ziye Jia$^{\ast}$, Qihui Wu$^{\ast}$$^{\dagger}$, Chao Dong$^{\ast}$, Zirui Zhuang$^{\ddagger}$, Huiling Hu$\textsuperscript{\textsection}$ and Qi Cai$\textsuperscript{\textsection}$\\
 }\IEEEauthorblockA{$^{\ast}$The Key Laboratory of Dynamic Cognitive System of Electromagnetic Spectrum Space, Ministry of Industry and Information Technology, Nanjing University of Aeronautics and Astronautics, 210016\\
$^{\ddagger}$State Key Laboratory of Networking and Switching Technology, Beijing University of Posts and Telecommunications, Beijing, China, 100876\\
$\textsuperscript{\textsection}$Middle-south Regional Air Traffic Management Bureau of CAAC, 510403\\
$^{\dagger}$Corresponding author, email: wuqihui@nuaa.edu.cn}}

\maketitle
\thispagestyle{empty}
\begin{abstract}
    With the increasing use of autonomous unmanned aerial vehicles (UAVs), it is critical to ensure that they are continuously tracked and controlled, especially when UAVs operate beyond the communication range of ground stations (GSs). Conventional surveillance methods for UAVs, such as satellite communications, ground mobile networks and radars are subject to high costs and latency. The automatic dependent surveillance-broadcast (ADS-B) emerges as a promising method to monitor UAVs, due to the advantages of real-time capabilities, easy deployment and affordable cost. Therefore, we employ the ADS-B for UAV trajectory tracking in this work. However, the inherent noise in the transmitted data poses an obstacle for precisely tracking UAVs. Hence, we propose the algorithm of recurrent neural network-enhanced interacting multiple model-Kalman filter (RNN-enhanced IMM-KF) for UAV trajectory filtering. Specifically, the algorithm utilizes the RNN to capture the maneuvering behavior of UAVs and the noise level in the ADS-B data. Moreover, accurate UAV tracking is achieved by adaptively adjusting the process noise matrix and observation noise matrix of IMM-KF with the assistance of the RNN. The proposed algorithm can facilitate GSs to make timely decisions during trajectory deviations of UAVs and improve the airspace safety. Finally, via comprehensive simulations, the total root mean square error of the proposed algorithm decreases by 28.56\%, compared to the traditional IMM-KF. 
\end{abstract}

\newcommand{\CLASSINPUTtoptextmargin}{0.8in}

\newcommand{\CLASSINPUTbottomtextmargin}{1in}

\section{Introduction}

\lettrine[lines=2]
U{nmanned} aerial vehicles (UAVs) draw a large amount of attentions 
due to leading economic development in the low-altitude field \cite{1}, \cite{jia1}. 
In civilian scenarios,  such as industrial inspection and goods delivery, 
the autonomous flight technology is widely applied to UAVs \cite{jia3}. 
In detail, UAVs follow the pre-planed flight path by the ground station (GS) to carry out the mission.
However, due to the autonomy of UAVs, 
real-time flight monitoring information can not be transmitted to GSs when UAVs exceed the communication range of GSs.
Therefore, it is challenging to monitor UAVs in real time to ensure flight safety.
In response to this challenge, the satellite communication, ground mobile network communication and radar surveillance emerge as viable options for monitoring flight trajectories. 
However, the aforementioned techniques are limited by their properties. For example, satellite communications 
have higher latency and are susceptible to adverse weather conditions \cite{3}.   
Ground mobile network communications may suffer from signal interference in densely populated areas and 
their coverage is affected by the distribution of GSs \cite{4}. 
In addition, radar deployment and construction are constrained by cost and terrain.
\textcolor{black}{Compared to the above methods, 
the automatic dependent surveillance-broadcast (ADS-B) 
presents an efficient aircraft monitoring solution with
low cost, low latency and wide coverage \cite{5}. Specifically, ADS-B consists of two 
components: ADS-B OUT and  ADS-B IN. ADS-B OUT can be installed on UAVs as a mobile unit to 
broadcast real-time flight status information, such as the position and speed. ADS-B IN is used on 
the ground to receive and decode the received data from ADS-B OUT \cite{6}. Such real-time information can assist GSs to make
timely decisions when anomalous behaviors occur, such as deviations from the flight path.}
\textcolor{black}{However, UAVs possess small size, high maneuverability and 
strong autonomous flight capabilities, imposing higher accuracy requirements 
on their flight trajectory monitoring \cite{7}. 
Moreover, there exists noise caused by signal transmission
interference, atmospheric disturbance and other factors during
data transmission process via ADS-B. It is intractable to make the
analysis and assessment of the true trajectory status of UAVs
due to the noise.}

\textcolor{black}{Considering the challenges posed by noise during data transmission via ADS-B, 
it becomes crucial to employ effective filtering techniques to reduce the impact of noise. In this context, the Kalman filter (KF) is employed as a key technology in various 
fields such as control, estimation and navigation, and is widely used in 
target tracking \cite{8}. 
Leveraging the dynamic model and the observation equation of the target system, 
KF obtains accurate state estimation.}
\textcolor{black}{
However, when tracking maneuvering targets, it is challenging due to the model mismatch, leading
to performance degradation or result divergence \cite{hong2022improvement}. 
In this context, \cite{10} introduces the interacting multiple model (IMM) algorithm, which is commonly 
used to track maneuvering targets.
By defining multiple dynamic models of target and switching among them 
according to probabilities, IMM enables continuous tracking. Furthermore, deep learning technologies 
show remarkable effectiveness in improving the performance of filters by 
learning information from 
large amounts of data \cite{jia2}. For example, \cite{11} proposes an adaptive tracking method based on long short-term memory (LSTM),} but the designed algorithm lacks interpretability of 
the model due to the abandonment of the traditional filtering framework. 
\cite{revach2022kalmannet} proposes a KalmanNet neural network (NN) based on gated recurrent unit cells, 
and combines with KF to achieve real-time state estimation.
However, these works do not consider the scenario of applying ADS-B to UAV trajectory tracking. 
Besides, the system model for UAVs in three-dimensional space is ignored.
% \textcolor{black}{
% Additionally, we employ a data-driven approach, utilizing a recurrent 
% neural network (RNN) to adaptively adjust the process noise and observation noise 
% matrices during the Kalman filtering process. By learning the dynamic 
% characteristics of the system from the data, we enable the IMM-KF to better 
% adapt to the nonlinear changes in the system, thereby enhancing the accuracy 
% and robustness of state estimation.}

\textcolor{black}{
    Therefore, in this paper, we emphasis on
    utilizing ADS-B as a method of transmitting UAV trajectory
    information to achieve real-time situation awareness of UAV flight. To filter the noise in ADS-B data, 
    we adopt an IMM-KF algorithm and define dynamic models corresponding to different flight modes 
    of maneuvering UAVs. To tackle issues from the ever-changing noise and uncertainties of accurate models, 
    we design a recurrent neural network-enhanced (RNN-enhanced) IMM-KF algorithm.
    By conducting simulation results compared to traditional IMM-KF, the superiority of the proposed algorithm 
    is verified in overcoming performance degradation caused by model 
    mismatch and inaccuracy. Also, the approach 
    realizes precise monitoring of UAV trajectories.
}

The rest of the paper is organized as follows. The system model and problem formulation are
presented in Section \ref{sec:System-Model}. Then, Section \ref{sec:algorithm} proposes the RNN-enhanced IMM-KF algorithm. Numerical results
are provided in Section \ref{sec:Simulation Result} and conclusions are drawn in Section \ref{sec:Conclusions}.

\section{System Model And Problem Formulation\label{sec:System-Model}}
\subsection{Trajectory and Observation Model Based on ADS-B}
\begin{figure}[t]
\centering
\includegraphics[scale=0.25]{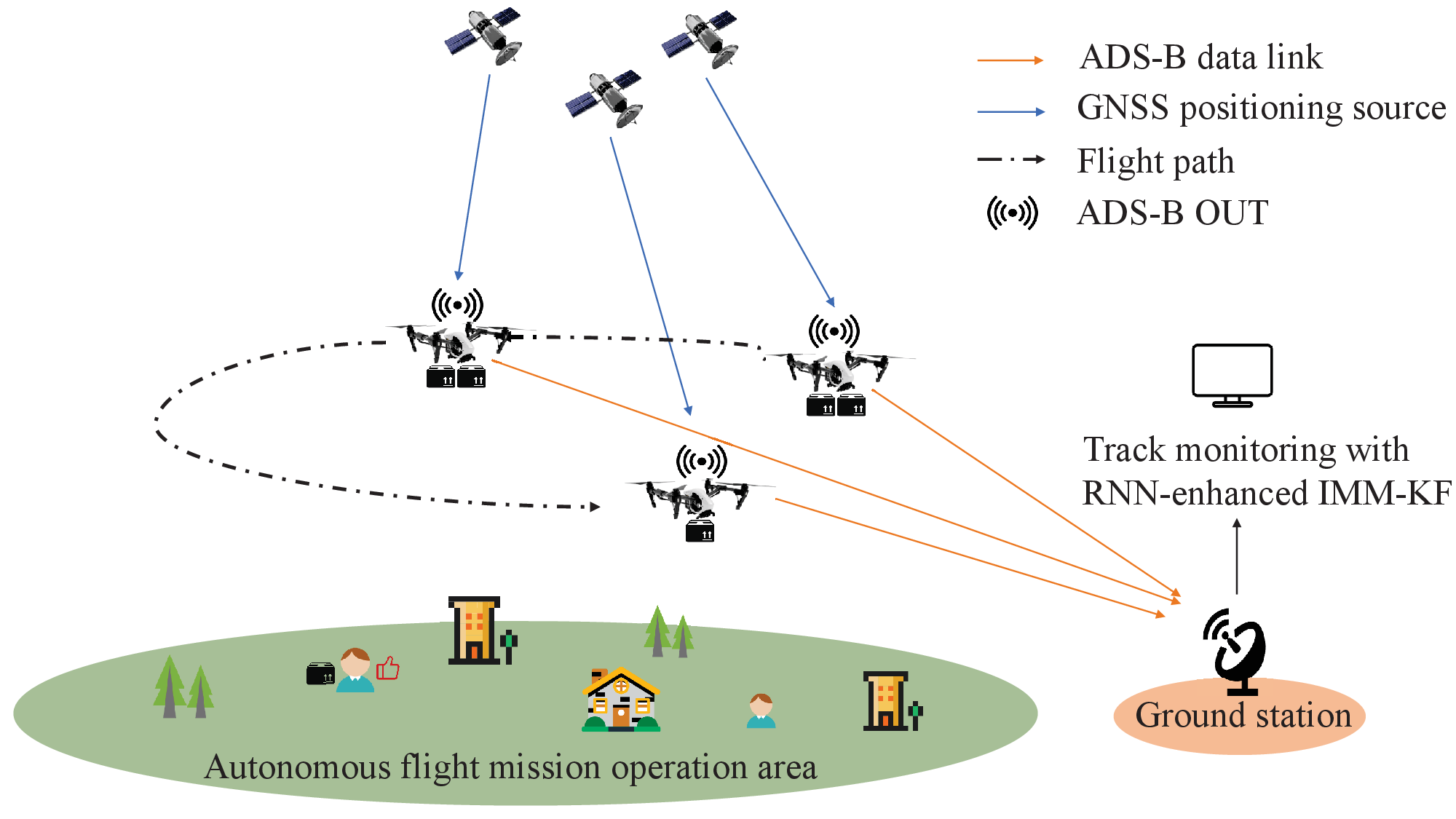}\caption{\label{fig:TEG}UAV trajectory tracking system with ADS-B information.}
% \vspace{-5mm}
\end{figure}

As depicted in Fig. \ref{fig:TEG}, a UAV equipped with ADS-B executes autonomous 
flight tasks in a specific operational area. 
The UAV transmits key state information data obtained from the on-board sensing devices to the GSs via the ADS-B. 
The state information of the UAV received by the GSs at discrete-time step $k \in \mathbb{T}$ is denoted as $S_k=\left\{\operatorname{lon}_k, \operatorname{lat}_k, 
\operatorname{alt}_k, v_k, \psi_k, \theta_k\right\}$. 
$\operatorname{lon}_k$, $\operatorname{lat}_k$, $\operatorname{alt}_k$, $v_k$, $\psi_k$ and $\theta_k$ represent the longitude, latitude, 
altitude, speed, heading angle, and pitch angle, respectively.
It is observed that the data transmitted by ADS-B provides 
comprehensive, real-time flight state information of the UAV.
For computational convenience and uniformity in data processing, we standardize the units and coordinate system of the observed data. 
In other words, the latitude and longitude information of the UAV, which is obtained from ADS-B and operated in the World Geodetic System 1984 coordinate system, is firstly mapped into 
the Cartesian coordinate system using the Gauss-Kruger projection \cite{13}. 
Furthermore, the velocity $v_k$ is decomposed into the velocity components $(v_{x_k}, v_{y_k}, v_{z_k})$ in the Cartesian coordinate system by heading angle $\psi_k$ and pitch angle $\theta_k$. 
According to the coordinate transformation and velocity decomposition, the final observation data 
from ADS-B at time instance $k$ is given as 
$\mathbf{o}_k=\left[x_k, y_k, z_k, v_{x_k}, v_{y_k}, v_{z_k}\right]^{\text{\tiny $T$}}$.
Additionally, the observation equation for the system is expressed as:
\begin{equation}
    \mathbf{o}_k=\mathbf{H} \cdot \mathbf{x}_k+\mathbf{v}_k, \quad \mathbf{v}_k \sim \mathcal{N}(0, \mathbf{R}), \quad \mathbf{o}_k \in \mathbb{R}^n,
    \end{equation}
    where $\mathbf{H}$ is the observation matrix linking the state $\mathbf{x}_k$ to the observed data, and $\mathbf{v}_k$ is 
    the Gaussian white noise with covariance matrix $\mathbf{R}$. In this scenario, the covariance matrix $\mathbf{R}$ 
      is expressed as $\mathbf{R}=diag\left(\sigma_x^2, \sigma_y^2, \sigma_z^2, \sigma_{v x}^2, \sigma_{v y}^2, \sigma_{v z}^2\right)$.
\vspace{-2mm}
\subsection{Dynamic System Model}\label{sec:model}
We consider the discrete-time state transition equation to describe the dynamic system of UAVs, denoted as 
\begin{equation}
    \mathbf{x}_k=\mathbf{F} \cdot \mathbf{x}_{k-1}+ \boldsymbol{\omega}_k, \quad \boldsymbol{\omega}_k \sim \mathcal{N}(0, \mathbf{Q}), \quad \mathbf{x}_k \in \mathbb{R}^m,
\end{equation} 
    where $\mathbf{F}$ is the state transition matrix and $\boldsymbol{\omega}_k$ represents Gaussian white noise with covariance matrix $\mathbf{Q}$.

    Considering that the flight patterns of UAVs are unable to be characterized accurately by a single model,  we define two types of 
    flight states: stable cruising and high-agility maneuvering, respectively. The corresponding two dynamic models are described as
    follows:
\subsubsection{\textcolor{black}{Constant Velocity (CV) model}}
The state of UAVs in stable cruising state scenario is characterized 
by the CV model. In detail, the state of UAVs is captured with the position 
$\left(x_k, y_k, z_k\right)$ and velocity $\left(v_{x k}, v_{y k}, v_{z k}\right)$. Accordingly, 
we derive the state vector $\mathbf{x}_k=\left[x_k, y_k, z_k, v_{x k}, v_{y k}, v_{z k}\right]^{\text{\tiny $T$}}$. 
The associated state transition matrix $\mathbf{F}_{C V}$ and process noise covariance matrix $\mathbf{Q}_{C V}$ are respectively given as:
\begin{equation}\label{eq:FCV}
    \mathbf{F}_{C V}=\left[\begin{array}{ll}
    \mathbf{I}_{3 \times 3} & T \mathbf{I}_{3 \times 3} \\
    0_{3 \times 3} & T \mathbf{I}_{3 \times 3}
    \end{array}\right],
    \end{equation}
and
\begin{equation}\label{eq:QCV}
    \mathbf{Q}_{C V}=\sigma_{v w}^2\left[\begin{array}{cc}\frac{T^3}{2} \mathbf{I}_{3 \times 3} & \frac{T^2}{2} \mathbf{I}_{3 \times 3} \\ \frac{T^2}{2} \mathbf{I}_{3 \times 3} & T \mathbf{I}_{3 \times 3}\end{array}\right],
\end{equation}
where $T$ represents the periodic time interval at which ADS-B OUT system broadcasts the state, and $\mathbf{I}$ denotes the identity matrix.
$\sigma_{v w}^2$ indicates the variance associated with the velocity of UAVs, which can also be set separately in the dimensions of $x$, $y$, and $z$.
In a physical sense, the matrix \( \mathbf{F}_{C V} \) indicates the UAVs move in a straight line with a uniform speed in all directions over time interval $T$.
In other words, UAVs are in stable cruising state. 
Additionally, the process noise matrix \( \mathbf{Q}_{C V} \) enables the filter to better cope with potential nonlinear motions of UAVs.
In practice, it improves the robustness of unpredictable changes in the velocity of UAVs that cannot be directly captured by a linear model.

\subsubsection{\textcolor{black}{Constant Jerk (CJ) model}} For high-agility maneuvers, CJ model can be used to represent the motion of UAVs. The state vector of UAVs is 
given by $\mathbf{x}_k=\left[x_k, y_k, z_k, v_{x k}, v_{y k}, v_{z k}, a_{x k}, a_{y k}, a_{z k}, j_{x k}, j_{y k}, j_{z k}\right]^{\text{\tiny $T$}}$, where $(a_{x k}, a_{y k}, a_{z k})$
represents the accelerations in directions of $x$, $y$ and $z$, respectively, and $(j_{x k}, j_{y k}, j_{z k})$
denotes the jerks, i.e., the change rate of accelerations. 
According to \cite{mehrotra1997jerk}, the corresponding state transition matrix $\mathbf{F}_{C J}$ and 
the process noise covariance matrix $\mathbf{Q}_{C J}$ are described as:
\renewcommand{\arraystretch}{1.1}
\begin{equation}\label{eq:FCJ}
    \mathbf{F}_{C J}=\left[\begin{array}{cccc} 
    \mathbf{I}_{3 \times 3} & T \mathbf{I}_{3 \times 3} & \frac{T^2}{2} \mathbf{I}_{3 \times 3} & \frac{T^3}{6} \mathbf{I}_{3 \times 3} \\
    0_{3 \times 3} & \mathbf{I}_{3 \times 3} & T \mathbf{I}_{3 \times 3} & \frac{T^2}{2} \mathbf{I}_{3 \times 3} \\
    0_{3 \times 3} & 0_{3 \times 3} & \mathbf{I}_{3 \times 3} & T \mathbf{I}_{3 \times 3} \\
    0_{3 \times 3} & 0_{3 \times 3} & 0_{3 \times 3} & \mathbf{I}_{3 \times 3}
    \end{array}\right],
    \end{equation}
    and
\renewcommand{\arraystretch}{1.7}
\begin{equation}\label{eq:QCJ}
    \mathbf{Q}_{C J}=\sigma_{j w}^2\left[\begin{array}{llll}
    \frac{T^7}{252} \mathbf{I}_{3 \times 3} & \frac{T^6}{72} \mathbf{I}_{3 \times 3} & \frac{T^5}{72} \mathbf{I}_{3 \times 3} & \frac{T^4}{24} \mathbf{I}_{3 \times 3} \\
    \frac{T^6}{72} \mathbf{I}_{3 \times 3} & \frac{T^5}{20} \mathbf{I}_{3 \times 3} & \frac{T^4}{8} \mathbf{I}_{3 \times 3} & \frac{T^3}{6} \mathbf{I}_{3 \times 3} \\
    \frac{T^5}{72} \mathbf{I}_{3 \times 3} & \frac{T^4}{8} \mathbf{I}_{3 \times 3} & \frac{T^3}{3} \mathbf{I}_{3 \times 3} & \frac{T^2}{2} \mathbf{I}_{3 \times 3} \\
    \frac{T^4}{24} \mathbf{I}_{3 \times 3} & \frac{T^3}{6} \mathbf{I}_{3 \times 3} & \frac{T^2}{2} \mathbf{I}_{3 \times 3} & \mathrm{\mathit{T}\mathbf{I}}_{3 \times 3}
    \end{array}\right].
    \end{equation}
Recall that $T$ is defined as the frequency at which the ADS-B OUT system broadcasts 
its state. $\sigma_{j w}^2$ denotes the variance with respect to the jerk of UAVs, 
reflecting the uncertainty of fast UAV maneuvers. Similarly, $\sigma_{j w}^2$ can be set on dimensions of $x$, $y$, and $z$.
The physical implication of the matrix $\mathbf{F}_{C J}$ is that UAVs move in all directions with the varying accelerations over a time interval $T$.
For example, $\mathbf{F}_{C J}$ applies to UAVs in situations involving swift obstacle avoidance or sharp turns.
Taking into account the uncertainty of the jerk estimation during the movement of UAVs, the robustness of the state estimation is further enhanced by matrix $\mathbf{Q}_{C J}$.
\vspace{-3mm}
\subsection{Problem Formulation}
\subsubsection{\textcolor{black}{Significance of $\mathbf{Q}$ and $\mathbf{R}$}}
In the field of dynamic traget tracking, robust algorithms are important. 
The IMM-KF emerges as a frontrunner for its trajectory tracking 
capabilities. In practice, numerous implementations leverage IMM-KF for effective dynamic 
target tracking. 
Among them, a key factor affecting the filter efficiency is the precise setting  
of the parameters, especially the process noise covariance matrix $\mathbf{Q}$ and the observation
noise covariance matrix $\mathbf{R}$. Specifically, the martices $\mathbf{Q}$ and $\mathbf{R}$ capture the uncertainties 
concerning dynamic models and the observation system, respectively. 
In detail, the matrices are employed to obtain the Kalman 
gain in KF, which determines the weight assigned to the observation in the update 
step. Additionally, $\mathbf{Q}$ plays an important role in IMM during the 
model transition process. A smaller $\mathbf{Q}$ value improves the accuracy of the KF, 
it may cause delay in the model transition 
phase, resulting in weaker timely tracking.

\subsubsection{Adaptive Tuning of $\mathbf{Q}$ and $\mathbf{R}$}
 Traditionally, in standard KF applications, $\mathbf{Q}$ and 
$\mathbf{R}$ are often derived from empirical knowledge and are set as 
fixed parameters. This approach may work well for 
systems with consistent and predictable dynamics. However, 
facing system environment changes and target motion nonlinearity, the static matrices lead to 
filter divergence and performance degradation, 
thus reducing the reliability of the tracking system.
Therefore, adaptive tuning the accurate parameters of the matrices is a primary 
challenge. It becomes tricky when domain knowledge is limited, such as partially known models and unknown statistical properties of noise.

\subsubsection{\textcolor{black}{Data-driven Adaptive Tuning}}
Considering the challenge of setting suitable $\mathbf{Q}$ and $\mathbf{R}$ in scenarios with limited 
domain knowledge, a data-driven approach emerges as 
a promising alternative. 
Based on the information learned by the NN from the data, the dynamic adjustment of 
$\mathbf{Q}$ and $\mathbf{R}$ can compensate for the inaccuracies of system modeling and ensure that the 
filter adapts to the noise characteristics. In this way, the traditional filtering 
method can be enhanced.

\subsubsection{\textcolor{black}{Implementation with RNNs}}
RNNs with their inherent capabilities to capture 
sequential patterns, present a compelling approach to the problem of adaptively tuning $\mathbf{Q}$ and $\mathbf{R}$. 
These networks can be trained in a supervised manner with groundtruth data. 
In the context of UAVs, high-precision sensor, such as real-time kinematic, 
can be deployed to obtain groundtruth datasets. 
As the filtering process continues, matrices $\mathbf{Q}$ and $\mathbf{R}$ are continuously updated 
with the help of RNN to ensure that IMM-KF operates under optimal settings.

In summary, while $\mathbf{Q}$ and $\mathbf{R}$ are fundamental parameters 
for KF, it is challenging to obtain their optimal values. 
The utilization of RNNs presents a promising solution. 
During the process of UAV trajectory filtering, the tracking errors 
are key indicators for evaluating the performance of different methods. 
Therefore, we resort to the root mean square error (RMSE), a measure to illustrate the magnitude of the state estimation error. The optimization problem is defined as follows:
\begin{equation}
    \footnotesize
    \begin{gathered}
    \min _{\Theta_{\mathrm{R}}, \Theta_{\mathrm{Q}_{C V}}, \Theta_{\mathrm{Q}_{C J}}} \sqrt{\frac{1}{3 N} \sum_{k=1}^N\left(\left(x_k-\hat{x}_k\right)^2+\left(y_k-\hat{y}_k\right)^2+\left(z_k-\hat{z}_k\right)^2\right)}, 
    \end{gathered}
    \normalsize
    \end{equation}
where $N$ is the length of the ADS-B sampled sequence in the
entire flight trajectory. $(\hat{x}_k, \hat{y}_k, \hat{z}_k)$ and  $(x_k, y_k, z_k)$ represent the 
three-dimensional position state estimation from the filter and the realistic state values at time $k$, respectively. 
$\Theta_{\mathrm{Q}_{C V}}=\left\{\sigma_{v w x}, \sigma_{v w y}, \sigma_{v w z}\right\}$, $\Theta_{\mathrm{Q}_{C J}}=\left\{\sigma_{j w x}, \sigma_{j w y}, \sigma_{j w z}\right\}$ and $\Theta_{\mathrm{R}}=\left\{\sigma_x, \sigma_y, \sigma_z, \sigma_{v x}, \sigma_{v y}, \sigma_{v z}\right\}$ 
represent the real-time matrix parameter sets to be estimated in the filtering algorithm.

In order to obtain the minimum RMSE across all dimensions, the filtering 
algorithm is expected to be equipped with the ability to 
adjust the optimal $\Theta_Q$ and $\Theta_R$ in real time.
In Section \ref{sec:algorithm}, we propose the
RNN-enhanced IMM-KF algorithm to deal
with the optimization problem.

\section{RNN-enhanced IMM-KF Algorithm \label{sec:algorithm}}
In this section, we present the RNN-enhanced IMM-KF algorithm. 
The diagram of the algorithm is depicted in Fig. \ref{fig:IMM}.
The right part is the KF with the specific RNN, which has the ability to adjust noise matrices dynamically, illustrated in Section \ref{sec:kf}.
On the left part of Fig. \ref{fig:IMM}, the IMM structure incorporates KF1 and KF2 with CV model and CJ model, respectively, illustrated in Section \ref{sec:IMM}.
The design of the specific RNN and the overall algorithmic workflow are demonstrated in Section \ref{sec:RNN}.
\vspace{-1mm}
\subsection{Kalman Filter with Dynamic Noise Matrices}\label{sec:kf}
In this work, we improve the performance of the IMM-KF by employing an specific RNN to dynamically 
generate time-varying noise matrices. Such matrices replace the process and observation noise 
matrices that are fixed in the prediction and the state update process of KF.
\subsubsection{\textcolor{black}{Prediction Phase}}
The prediction phase predicts the current prior state estimation and error covariance based on previous state estimations. The formulations are given as follows: 
\begin{equation}
    \mathbf{\hat{x}}_{k \mid k-1}=\mathbf{F} \mathbf{\hat{x}}_{k-1 \mid k-1},
\end{equation}
and
\begin{equation}
    \mathbf{P}_{k \mid k-1}=\mathbf{F} \mathbf{P}_{k-1 \mid k-1} \mathbf{F}^T+\mathbf{Q}_k,
\end{equation}
where $\mathbf{F}$ and $\mathbf{Q}_k$ denote the state transition matrix and the process noise matrix generated by the RNN, respectively. 
% In ADS-B-based UAV trajectory tracking, 
% direct control inputs to the target are not observed. 
% Therefore, the term $\mathbf{B}_k \mathbf{u}_k$ is omitted from estimation of the dynamic model.
\begin{figure}[t]
    \hspace{-3.8mm}
    \centering
    \includegraphics[scale=0.27]{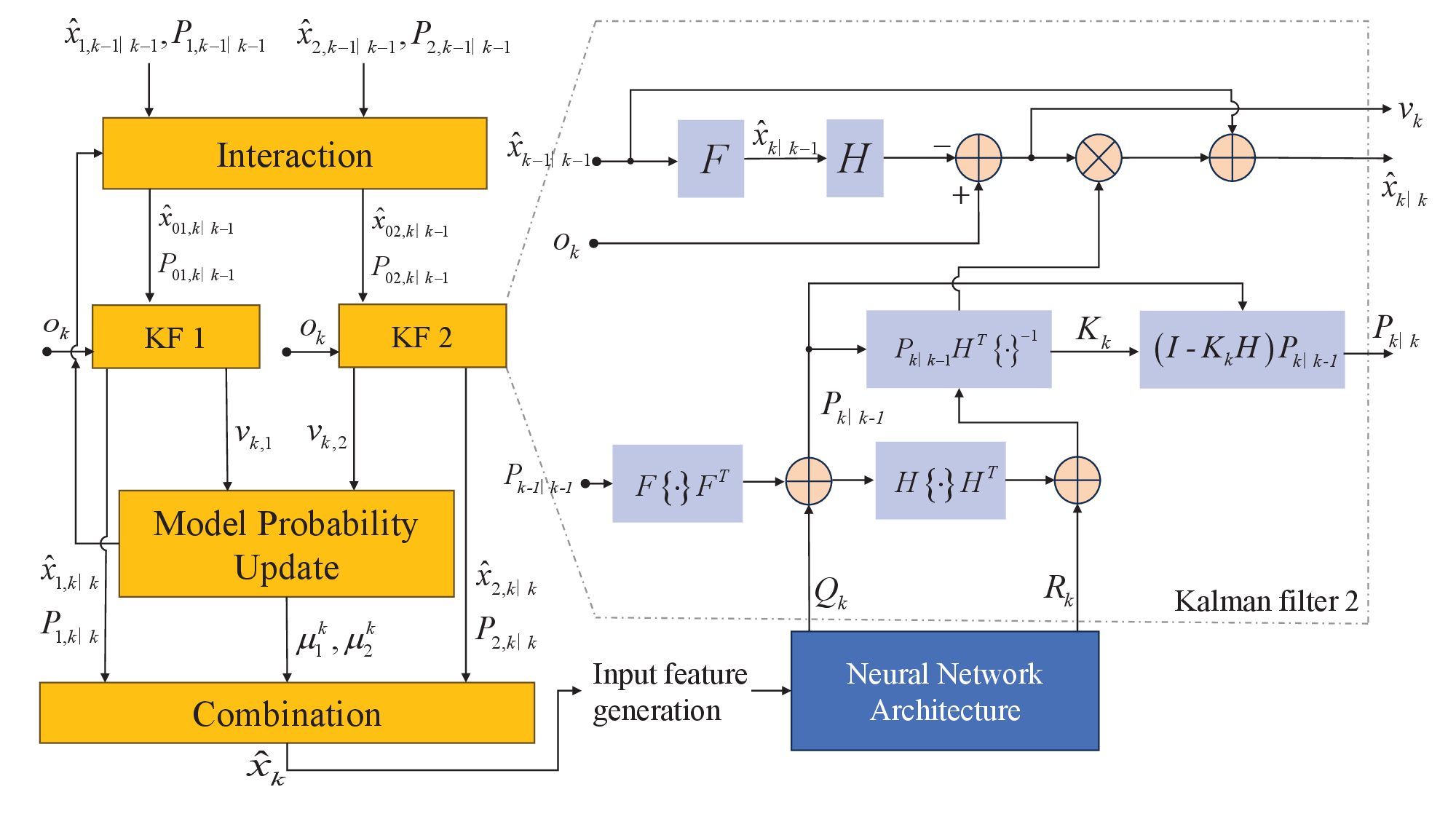}\caption{\label{fig:IMM}RNN-enhanced IMM-KF block diagram.}
    \vspace{-5mm}
\end{figure}

\subsubsection{\textcolor{black}{Update Phase}}
In the update phase, the updated state estimation and error covariance are computed as:
\begin{equation}
    \mathbf{\hat{x}}_{k \mid k}=\mathbf{\hat{x}}_{k \mid k-1}+\mathbf{K}_k\left(\mathbf{o}_k-\mathbf{H} \mathbf{\hat{x}}_{k \mid k-1}\right),
\end{equation}
and
\begin{equation}
    \mathbf{P}_{k \mid k}=\left(\mathbf{I}-\mathbf{K}_k \mathbf{H}\right) \mathbf{P}_{k \mid k-1},
\end{equation}
where the Kalman gain $\mathbf{K}_k$ is derived from prior error 
covariance $\mathbf{P}_{k \mid k-1}$, observation matrix $\mathbf{H}$, and observation noise matrix $\mathbf{R}_k$ generated by the RNN, given by:
\begin{equation}
    \mathbf{K}_k=\mathbf{P}_{k \mid k-1} \mathbf{H}^T\left(\mathbf{H} \mathbf{P}_{k \mid k-1} \mathbf{H}^T+\mathbf{R}_k\right)^{-1}.
\end{equation}
\vspace{-2mm}
\subsection{IMM for Fusion of CV and CJ Models}\label{sec:IMM}
The IMM algorithm is used to combine the two dynamic models defined in Section \ref{sec:model}. As depicted in Fig. \ref{fig:IMM}, the CV and CJ models are used for KF1 and KF2, respectively.
 Specifically, the IMM consists of four primary components:
\subsubsection{\textcolor{black}{Interaction}}
In the interaction phase, the mixing probability $\mu_{i j}^k$ is 
introduced to prepare the initial states for the filters corresponding 
to different models. In detail,
\vspace{-1mm}
\begin{equation}
    \mu_{i j}^k=\frac{\mu_i^{k-1} \lambda_{i j}}{\sum_{l=1}^2 \mu_l^{k-1} \lambda_{l j}}, \forall i, j \in\{1,2\},
    \end{equation}
where $\lambda_{i j}$ and $\mu_j^{k-1}$ represent the probability of model transition and the posterior model, respectively. $i$ and $j$ denote the dynamic system models in IMM-KF. The state estimation and 
error covariance are the input to the KF corresponding to model $j$, computed as:
\begin{equation}
    \mathbf{\hat{x}}_{0 j, k \mid k-1}=\sum_{i=1}^2 \mu_{i j}^k \mathbf{\hat{x}}_{i, k-1 \mid k-1},
\end{equation}
\vspace{-3mm}
and
\begin{align}
    \mathbf{P}_{0 j, k \mid k-1} &= \sum_{i=1}^2 \mu_{i j}^k \Big( \mathbf{P}_{i, k-1 \mid k-1} \nonumber + \left(\mathbf{\hat{x}}_{i, k-1 \mid k-1}-\mathbf{\hat{x}}_{0 j, k \mid k-1}\right) \nonumber \\
    &\quad \times \left(\mathbf{\hat{x}}_{i, k-1 \mid k-1}-\mathbf{\hat{x}}_{0 j, k \mid k-1}\right)^T \Big).
\end{align}
\subsubsection{\textcolor{black}{Filtering}}
Two KFs operate in parallel, executing the state prediction and 
update process, as described in Section \ref{sec:kf}.
\subsubsection{\textcolor{black}{Model Probability Update}}
$\mu_j^k$ 
represents the probability that the target is in model $j$ at time $k$. $\mu_j^k$ serves as the basis for the weights of each model considered in the 
final estimation result, i.e., 
\begin{equation}
    \mu_j^k=\frac{\mathcal{L}_j\left(\mathbf{o}_k\right) \sum_{i=1}^2 \lambda_{i j} \mu_i^{k-1}}{\sum_{l=1}^2 \mathcal{L}_l\left(\mathbf{o}_k\right) \sum_{i=1}^2 \lambda_{i l} \mu_i^{k-1}},
    \end{equation}
where the likelihood fucntion $\mathcal{L}_j$ for model $j$ at time $k$ is for measuring
the similarity between the predicted observation of the model and the actual observed data \cite{10120631}.
\subsubsection{\textcolor{black}{Combination}}
The combination step in the IMM algorithm combines the outputs from 
multiple models to produce a unified state estimation, denoted as
\begin{equation}
    \mathbf{\hat{x}}_k=\sum_{i=1}^2 \mu_i^k \mathbf{\hat{x}}_{i, k \mid k}.
    \end{equation}
\subsection{Neural Network Design for IMM-KF Enhancement}\label{sec:RNN}
\begin{figure}[t]
    \vspace{-2mm}
    \hspace{-7mm}
    \centering
    \includegraphics[scale=0.27]{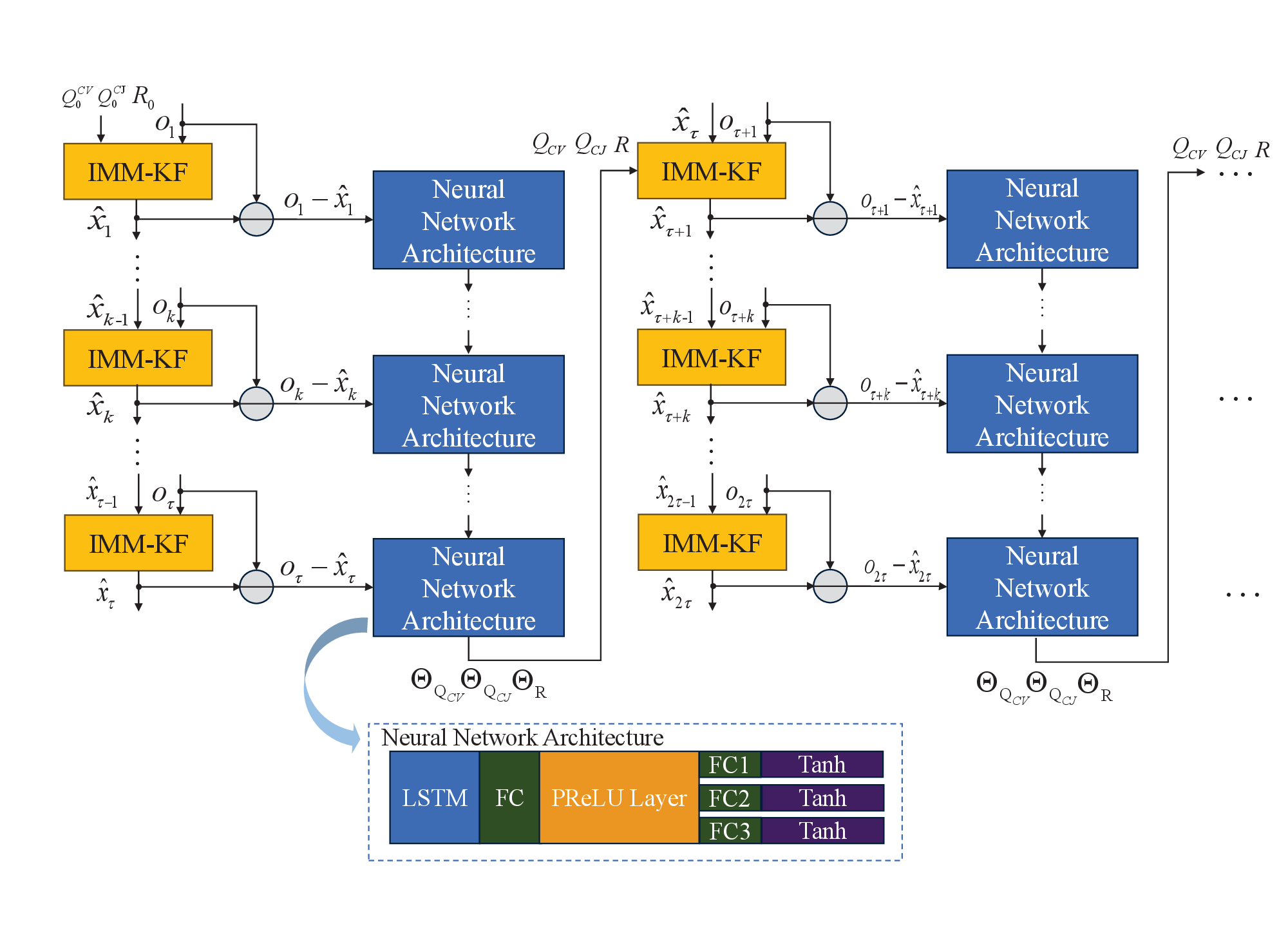}\vspace{-8mm}\caption{RNN-enhanced IMM-KF flowchart.}\label{fig:RNN}
    \vspace{-5mm}
\end{figure}

It is significant for NN to store and utilize multi-step 
historical data due to the handling of time series data involved in the filtering process.
Hence, we consider employing 
the NN architecture, depicted 
in Fig. \ref{fig:RNN}, to obtain the matrix parameters essential for the IMM-KF filtering process.
\subsubsection{\textcolor{black}{Architecture Design}}
Specifically, we adopt the LSTM cell to capture the historical 
sequence features. Subsequently, a fully connected (FC) layer is employed to map the 
hidden state of the LSTM cell to a fixed output dimension. Additionally, a 
parametric rectified linear unit (PReLU) activation function is introduced to incorporate non-linearity.
Finally, three individual FC layers, each followed by a $tanh$ 
activation function, are employed to output the estimated parameters $\Theta_{\mathrm{Q}_{C V}}$, $ \Theta_{\mathrm{Q}_{C J}}$ and $\Theta_{\mathrm{R}}$.
\subsubsection{\textcolor{black}{Input Feature and Algorithm Workflow}}
The designed NN is enabled to learn noise matrix parameters from data, 
which is used in the process of IMM-KF filtering. The 
workflow of the RNN-enhanced IMM-KF algorithm is illustrated in Fig. \ref{fig:RNN}. 
We employ a sliding window to realize the parameter estimation for adjusting and updating the noise matrices at a specific time interval. 
As shown in Fig. \ref{fig:RNN}, the length of the sliding window is $\tau$. For the sequence 
in the first time window, i.e., the received UAV flight trajectory 
sequence for the first $\tau$ time lengths, three noise matrices 
$\mathbf{Q}_0^{C V}$, $\mathbf{Q}_0^{C J}$ and $\mathbf{R}_0$ are initialized with empirical knowledge. 
Then, we conduct IMM-KF filtering to obtain the state estimation with length of $\tau$.
We choose the difference between the observation 
value at each time step $k \in \mathbb{T}$ and the corresponding state estimation value, 
and then provide it as the input feature to the NN, denoted by $\mathbf{\delta}_k=\mathbf{o}_k-\hat{\mathbf{x}}_k$. In detail, 
$\mathbf{\delta}_k$ contains the possible model mismatch information in 
the IMM-KF filtering process. Since IMM-KF is a multi-model interactive 
filter, it considers different models and corresponding weights to give the 
 final state estimation. Leveraging the mismatch information can help IMM-KF to correct 
 the model probability in time. Besides, $\mathbf{\delta}_k$ contains the noise level information of observation data. 
 The information can assist IMM-KF to adapt to different levels of observation noise.

In the subsequent time window, IMM-KF utilizes the noise matrix 
parameters $\Theta_{\mathrm{Q}_{C V}}$, $ \Theta_{\mathrm{Q}_{C J}}$ and $\Theta_{\mathrm{R}}$ 
to generate the noise matrices \( \mathbf{Q}_{C V} \), \( \mathbf{Q}_{C J} \) and $\mathbf{R}$, to complete subsequent filtering.
 \subsubsection{\textcolor{black}{Training Algorithm}}Our training algorithm adopts a supervised learning approach, 
 in which we use high-precision trajectories as label data. 
 In such way, the end-to-end training NN is derived. The advantage of this approach 
 is that it is not necessary to provide the real values of the outside parameters related to noise matrices. Instead, 
 it enables NN to automatically learn these parameters from the labeled data. 
 Additionally, the proposed method is 
 applicable to datasets with different trajectory lengths. The length of the time window is utilized as the time interval for conducting the backpropagation, 
 and the mean absolute error (MAE) with $L_1$ regularization is chosen as the loss function, since it is 
 more robust to tolerate outliers in the data than RMSE.

\section{Simulation Result\label{sec:Simulation Result}}
We use the fixed-wing UAV parser in MATLAB Toolbox to generate a dataset
containing 10 UAV trajectories, in which the simulated ADS-B sampling 
frequency is 1Hz, and the observation trajectory data includes the three-dimensional 
position and speed of the UAV at each sampling moment. The NN 
training parameters are set as: the hidden layer size of LSTM is 128, the sliding 
time window length is 3, training epoch is 15. Additionally, there are 7 trajectories used for training, 2 for validation and 1 for testing.
\begin{figure}[t]
    \centering
    \begin{minipage}{0.5\textwidth}
      \includegraphics[width=0.8\linewidth]{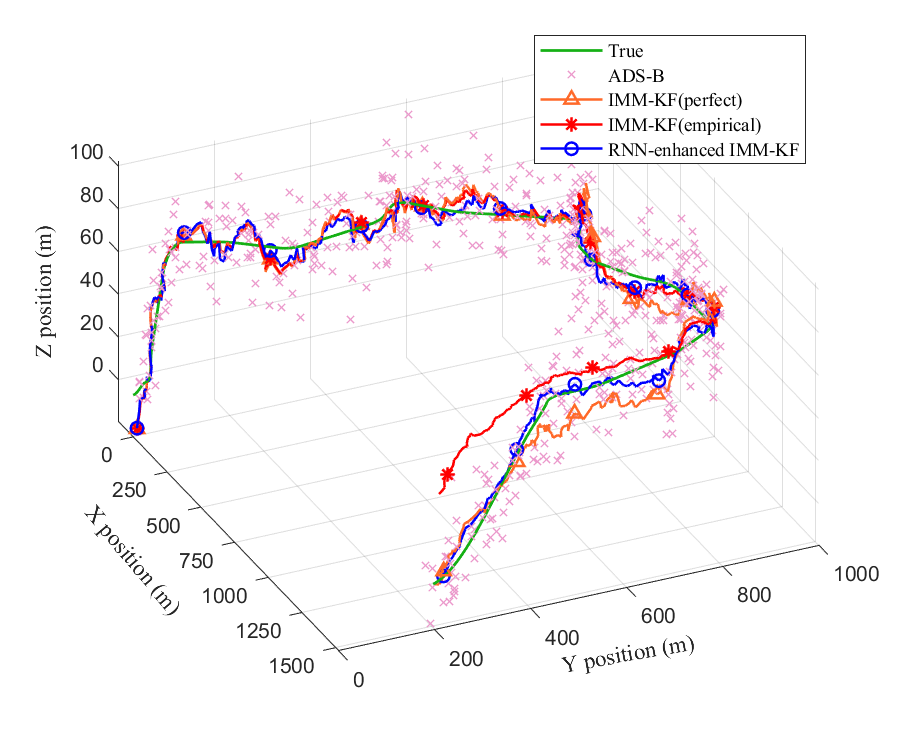}
      \caption{Tracking diagram of RNN-enhanced IMM-KF \textit{versus} IMM-KF.}
      \label{fig:traj1} 
    \end{minipage}
\end{figure}
Fig. \ref{fig:traj1} shows UAV tracking trajectories using RNN-enhanced IMM-KF and IMM-KF under different matrix parameter settings. 
Among them, IMM-KF (empirical) is the trajectory result obtained by directly 
using the noise matrix parameters of IMM-KF with good tracking performance 
on other trajectories and filtering. It is obvious that severe filter divergence occurs.
The IMM-KF (perfect) is the filtering result when the optimal noise matrix 
setting is obtained via reverse verification after all time data 
of the current trajectory sequence is obtained. It is observed that the reasonable 
configuration of the noise matrices $\mathbf{Q}_{C V}$, $\mathbf{Q}_{C J}$ and $\mathbf{R}$ has a positive 
impact on the filtering performance of IMM-KF. RNN-enhanced IMM-KF 
consistently maintains excellent tracking performance by continuously 
adjusting the noise matrix parameters.
\begin{figure}[htbp]
    \centering
    \begin{minipage}{\linewidth}
        \centering
        \subfloat[3D MAE at each time step.]{
            \includegraphics[width=0.8\linewidth]{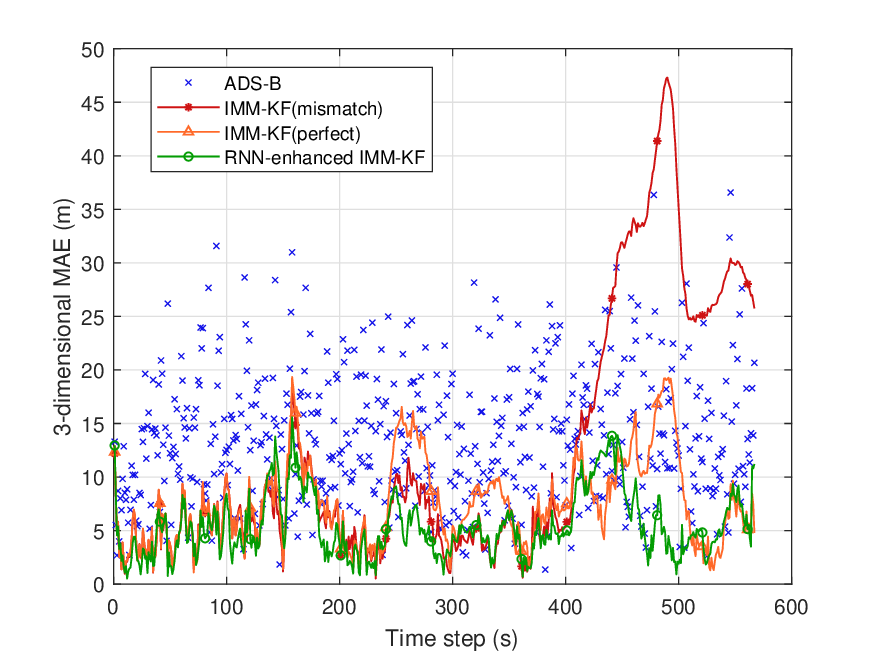}
            \label{fig:subfig1}
        }
        \vfill
        \subfloat[Comparison of RMSE across different dimensions.]{
        \includegraphics[width=0.8\linewidth]{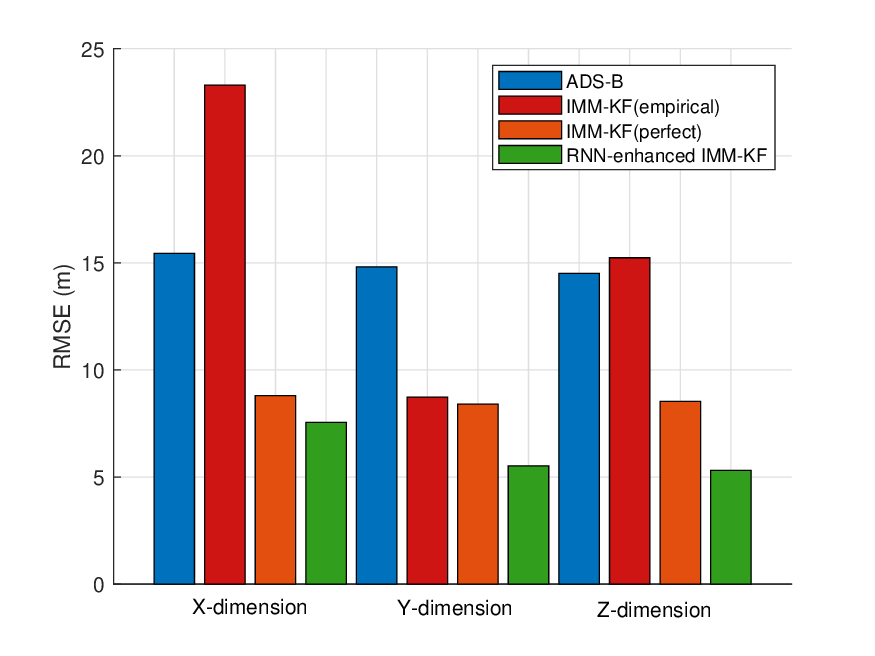}
            \label{fig:subfig2}
        }
        \caption{Error performance of RNN-enhanced IMM-KF \textit{versus} IMM-KF.}
        \label{fig:performance}
    \end{minipage}
\end{figure}

Fig. \ref{fig:performance} depicts the error performance of RNN-enhanced IMM-KF and IMM-KF algorithms. 
In detail, Fig. \ref{fig:subfig1} shows the three-dimensional MAE at each time step of the 
estimated values obtained using different filtering methods. It is obvious that 
within the first 400s, all three filtering methods maintain good target tracking 
and filtering performance. However, around 400s, the UAV performs a large degree of 
turning maneuvers during flight. Due to the incorrect setting of noise matrix parameters, 
the process of matching and calculation of correct model weights of IMM-KF (empirical) are 
affected, which leads to result divergence. By observing 
IMM-KF (perfect) and RNN-enhanced, there exists no filtering divergence in both, but the 
MAE performance of the proposed algorithm is better. It is explained that 
we set different levels of observation noise according to the
distance between the UAV and GS in the simulation dataset generation stage. Then, we can simulate 
and verify the filtering performance in changing environments. Obviously, 
RNN-enhanced IMM-KF can better adapt to changes in the environment and achieve 
better filtering performance. Fig. \ref{fig:subfig2} shows the RMSE among different filtering 
methods in different dimensions. The minimum RMSE achieved by the RNN-enhanced 
IMM-KF algorithm in all three spatial dimensions demonstrates its performance.
Compared with the original ADS-B data and IMM-KF (perfect) on the test trajectory, the total RMSE of the proposed algorithm decrease 
by 58.98\% and 28.56\%, respectively.
\section{Conclusions\label{sec:Conclusions}}
In this paper, we use ADS-B technology as a 
method of UAV target tracking. Considering the impact of receiving 
a large amount of ADS-B signal noise on tracking 
performance, we design the RNN-enhanced IMM-KF algorithm 
to execute real-time filtering of the target trajectory. The 
simulation results effectively verify the significance of the proposed 
algorithm in improving filtering performance and overcoming 
filtering divergence in UAV target tracking scenarios. This work can help 
improve the safety of UAV flight as well as airspace management.
\textcolor{black}{\bibliographystyle{IEEEtran}
\bibliography{ref}}
\end{document}